\documentclass[12pt,preprint]{aastex}
\begin{document}

\title{Quaking neutron star deriving radiative power
of oscillating magneto-dipole emission from energy of Alfv\'en seismic vibrations}

\author{S.I. Bastrukov\altaffilmark{1,2}, Molodtsova\altaffilmark{2}, J.W. Yu\altaffilmark{1} and  R.X.
 Xu\altaffilmark{1}}

\altaffiltext{1}{State Key Laboratory of Nuclear Physics and Technology,
School of Physics, \\
Peking University, Beijing 100871, China}

\altaffiltext{2}{Joint Institute for Nuclear Research, 141980 Dubna, Russia}

\begin{abstract}
 It is shown that monotonic depletion of the magnetic field pressure in  a quaking neutron
 star undergoing Lorentz-force-driven torsional seismic vibrations
 about axis of its dipole magnetic moment is accompanied by the loss of vibration
 energy of the star that causes its vibration period to lengthen at a rate proportional to
 the rate of  magnetic field decay.  Highlighted is the magnetic-field-decay induced
 conversion of the energy of differentially rotational Alfv\'en vibrations into the energy
 of oscillating magneto-dipole radiation. A set of representative examples
 of magnetic field decay illustrating the vibration energy powered emission with
 elongating periods produced by quaking neutron star are
 considered and discussed in the context of theory of magnetars.
 \end{abstract}

\section{Introduction}
 Most, if not all,  reported up to now computations of frequency spectra of poloidal
 and toroidal Alfv\'en vibration modes in pulsars and magnetars rest
 on tacitly adopted assumption about constant-in-time magnetic field in
 which a perfectly conducting neutron star matter undergoes Lorentz-force-driven
 oscillations (e.g. Carroll et al. 1986, Bastrukov et al. 1997, 1999, Lee 2007, 2008,
 Shaisultanov, Eichler 2009, Sotani et al. 2009; see, also, references therein). A special place in the study of the above Alfv\'en
 modes of pure shear magneto-mechanical seismic vibrations ($a$-modes) occupies a homogeneous model of a solid star with the uniform density
 $\rho$ and frozen-in poloidal static magnetic field of both homogeneous and
 inhomogeneous internal and dipolar external configuration.
 This field can be conveniently represented in the form
  \begin{eqnarray}
  \label{e1.1}
 {\bf B}({\bf r})=B\,{\bf b}({\bf r}),\quad {\bf b}({\bf r})=[b_r({\bf r})\neq 0,
  b_\theta({\bf r})\neq 0, b_\phi({\bf r})=0],\quad B={\rm constant},
  \end{eqnarray}
   where $B$ is the field intensity [in Gauss] and
   ${\bf b}({\bf r})$ stands for the dimensionless vector-function of spatial
   distribution of the field.  In particular, the above form of ${\bf B}({\bf r})$ has
   been
   utilized in recent works (Bastrukov et al. 2009a, 2009b, 2009c) in which the
   discrete spectra of frequencies of node-free
   torsional Alfv\'en oscillations has been computed in analytic form on the basis of
   equations of linear magneto-solid mechanics (Bastrukov et al. 2009a)
 \begin{eqnarray}
  \label{e1.1a}
 && \rho{\ddot {\bf u}}=\frac{1}{c}
 [\delta {\bf j}\times {\bf B}],\quad \nabla\cdot {\bf u}=0,\\
  \label{e1.1b}
 && \delta {\bf j}=\frac{c}{4\pi}[\nabla\times \delta {\bf B}],
 \quad \delta {\bf B}=\nabla\times [{\bf u}
  \times {\bf B}],\quad \nabla\cdot {\bf B}=0.
  \end{eqnarray}
   Here ${\bf u}={\bf u}({\bf r},t)$ is the field of material displacement, the
   fundamental dynamical variable of solid-mechanical theory of elastically deformable
   (non-flowing) material continua.  These equations describe  Lorentz-force-driven
    non-compressional vibrations of a perfectly conducting elastic matter of a solid
    star
    with Amp\'ere form of  fluctuating current density $\delta {\bf j}$. Equation for
    $\delta
    {\bf B}$ describing coupling between fluctuating field of
    material displacements ${\bf u}$
    and background magnetic field ${\bf B}$ pervading stellar material is the
    mathematical form of Alfv\'en
    theorem about frozen-in lines of magnetic field in perfectly conducting
    matter\footnote{Equation (\ref{e1.1a}) can be represented in the following
    equivalent tensor form (Bastrukov et al. 2009a),
    ${\ddot u}_{i}=\nabla_k \tau_{ik}$, where stress tensor
    $\tau_{ik}=(1/4\pi)[B_i\delta B_k+B_k\delta B_i-B_j\delta B_j\delta_{ik}]$ is
    the Maxwellian tensor of  magnetic field stresses. This last equation for $u_i$
    is identical in appearance to  canonical equation of solid-mechanics ${\ddot  u}_{i}=\nabla_k \sigma_{ik}$ where
    $\sigma_{ik}=2\mu\,u_{ik}+[\kappa-(2/3)\mu]\,u_{jj}\delta_{ik}$ is the
    Hookean stress tensor and $u_{ik}=(1/2)
    [\nabla_{i}\,u_k+\nabla_k\,u_i]$ is the tensor of shear deformations in an
    isotropic elastic continuous matter with shear
    modulus $\mu$ and bulk modulus
    $\kappa$. The adopted for Alfv\'en vibrations of compact solid stars
    terminology [which is
    not new of course, e.g., Moon (1984)], namely,
    magneto-solid mechanics  and solid-magnetics, is used as a solid-mechanical
    counterpart  of well-known terms like magneto-fluid mechanics,
    magnetohydrodynamics (MHD) and
    hydromagnetics, $\rho \delta {\dot {\bf v}}=(1/c)
 [\delta {\bf j}\times {\bf B}],\quad \delta {\bf j}=(c/4\pi)[\nabla\times \delta {\bf B}],\quad \delta {\dot {\bf B}}=\nabla\times [\delta {\bf v}
  \times {\bf B}]$, describing flowing magento-active plasma in terms of the velocity 
  of fluctuating flow $\delta {\bf v}={\dot {\bf u}}$  and fluctuating magnetic field ${\delta {\bf B}}$. 
   The MHD approach is normally utilized in astrophysics of main-sequence (MS) liquid stars. The advantage and virtue of the Rayleigh energy method 
 of computing frequencies of node-free Alfv\'en vibrations (Bastrukov et al. 2009a, 2009b, 2009c, Molodtsova et al. 2010) is that it 
    can too be applied to the above MHD equations 
    underlying  the study of $a$-modes in liquid MS-stars such, for instance,  as 
   rapidly oscillating Ap (roAp) stars, chemically peculiar magnetic
stars exhibiting high-frequency oscillations that have been and still remain 
the subject of extensive investigation (e.g., Ledoux and Walraven 1958, Rincon and Rieutord 2003).}.
    As for the general asteroseismology of compact objects is concerned, the above
    equations  seems to be appropriate not only for neutron stars but also white
    dwarfs (Molodtsova et al.
   2010) and quark stars (Heyvaerts et al. 2009). The superdense material of these
   latter yet hypothetical compact stars is too
   expected to be in solid state (Xu 2003, 2009), that is, in the aggregate state
   possessing mechanical property of shear elasticity.

    Inserting (\ref{e1.1a}) in (\ref{e1.1b})
  the former equation of solid-magnetics takes the form
    \begin{eqnarray}
  \label{e1.2}
 && \rho\, {\ddot {\bf u}}({\bf r},t)=\frac{1}{4\pi}
 [\nabla\times[\nabla\times [{\bf u}({\bf r},t)\times {\bf B}({\bf r})]]]\times {\bf B}
 ({\bf r}).
  \end{eqnarray}
  The obtained equation serves as a basis of our further analysis. The studied in
  above works vibration regime is of some interest in that the rate
  of differentially rotational material displacements
   \begin{eqnarray}
   \label{e1.3}
 && {\dot {\bf u}}({\bf r},t)=
  [\mbox{\boldmath $\omega$}({\bf r},t)\times {\bf r}],\quad
  \mbox{\boldmath $\omega$}({\bf r},t)=[\nabla\chi({\bf r})]\,{\dot\alpha}(t), \\
   \label{e1.4}
  && \nabla^2\chi({\bf
  r})=0,\quad \chi({\bf r})=f_\ell\,P_\ell(\cos\theta), \quad f_\ell(r)=A_\ell\, r^{\ell}
 \end{eqnarray}
  has one and the same form as in torsion elastic mode of node-free
  vibrations under the action of Hooke's force of mechanical 
  shear stresses\footnote{It may be worth noting that in standing-wave regime
  of elastic-force-driven differentially rotational vibrations, the toroidal 
  field of material displacements is described one and the same equation 
  (\ref{e1.3}), but function $\chi=A_\ell j_\ell(kr) P_\ell(\cos\theta)$, is the 
  solution of Helmholtz equation $\nabla^2\chi+k^2\chi=0$ where 
  $k^2=\omega^2/c^2_t$ and $c_t^2=\mu/\rho$ is the speed of shear wave of 
  material displacements in solid stellar matter of density $\rho$ and whose shear 
  modulus is $\mu$. The discrete frequency spectrum $\omega(\ell)=2\pi\nu(\ell)$
  of elastic torsional vibration of multipole degree $\ell$ is computed 
  from from boundary condition $n_k\sigma_{ik}=0$ resulting in 
  dispersion equation, $dj_\ell(z)/dz=j_\ell(z)/z$, where $z=kR$. 
  The characteristic feature of the long-wave regime of node-free 
  vibrations, $z<<1$, with the toroidal field of material displacements 
  (\ref{e1.3})-(\ref{e1.4}) is the absence of dipole overtone in the spectrum of 
  discrete frequencies (Bastrukov 2007a, 2007b): $\nu_e(\ell)=\nu_e[(2\ell+3)
  (\ell-1)]^{1/2}$, with basic frequency of shear elastic vibrations,  
  $\nu_e=c_t/R$, where $R$ stands for the star radius. 
  In the non-studied before  regime of short-wave length limit of torsional  elastic
  vibrations (the limit $z>>1$ taken in the exact dispersion equation for dipole 
  overtone $dj_1(z)/dz=j_1(z)/z$) leads to the spectrum of standing-wave dipole 
  vibrations of the following analytic form $\nu_e(\ell=1)=\nu_e(p+1)$, where $p=0,1,2,3...$ is the 
  node number of short-wavelength vibration regime.}
  (Bastrukov et al. 1999, 2007, 2008).
  Hereafter $P_\ell(\cos\theta)$ stands for
  Legendre polynomial of degree $\ell$ specifying the overtone of $a$-mode. The
  time-dependent amplitude $\alpha(t)$ describes temporal evolution of above
  vibrations; the governing equation for $\alpha(t)$ is obtained form equation
  (\ref{e1.2}).  The prime purpose of above works
  was to get some insight into difference between spectra of discrete frequencies
  of toroidal $a$-modes in neutron star models having one and the same
  mass  $M$ and radius $R$, but different shapes of constant-in-time poloidal
  magnetic fields. By use of the energy method, it was found that each specific
  form of spatial configuration
  of static magnetic field  about axis of which the neutron star matter undergoes
  nodeless torsional oscillations is uniquely reflected in the discrete frequency spectra
  by form of dependence of frequency upon overtone $\ell$ of nodeless vibration.

\begin{figure*}[ht]
\begin{center}
 \includegraphics[width=10cm]{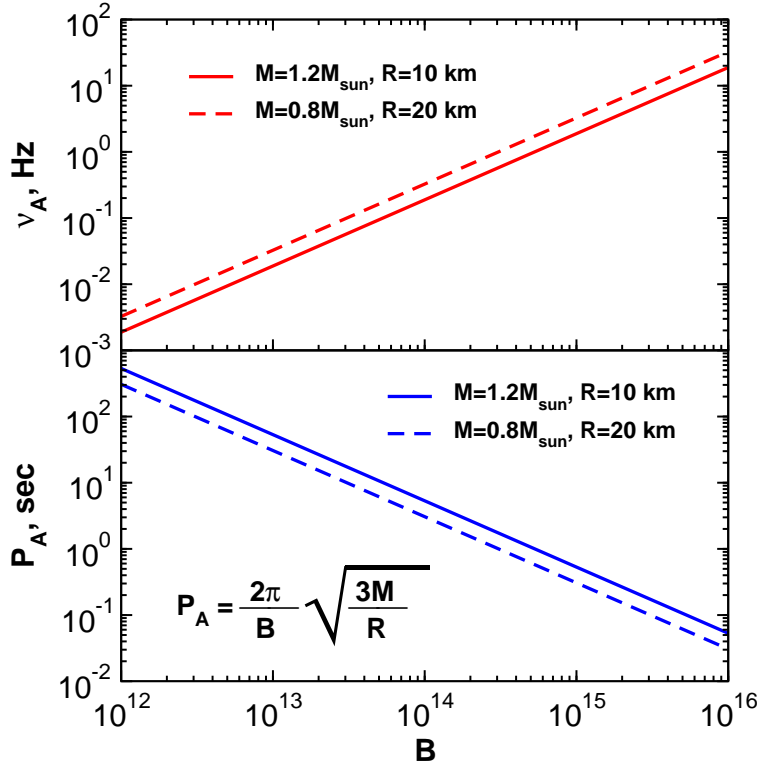}
 \end{center}
  \caption{
  The basic frequency and period of global Alfv\'en oscillations, equations (\ref{e1.6}), as functions of magnetic field intensity in the neutron star models with indicated mass and radius.}
  \label{fig:fr-period}
\end{figure*}

  The assumption about constant in time undisturbed magnetic field
  means that the internal magnetic field pressure, $P_B$, the velocity $v_A$ of
  Alfv\'en wave
  in the star bulk (e.g. Lee 2007)
   \begin{eqnarray}
  \label{e1.5}
   P_B=\frac{B^2}{8\pi}, \quad v_A=\sqrt{\frac{2P_B}{\rho}}
  \end{eqnarray}
  and, hence, the frequency $\nu_A=\omega_A/2\pi$
  (where $\omega_A=v_A/R$) and the period
  $P_A=\nu_A^{-1}$ of global Alfv\'en oscillations
  \begin{eqnarray}
  \label{e1.6}
 \nu_A=\frac{B}{2\pi}\sqrt{\frac{R}{3M}},\quad
 P_A=\frac{2\pi}{B}\sqrt{\frac{3M}{R}}
\end{eqnarray}
 remain constant in the process of vibrations whose amplitude $\alpha(t)$
 subjects to standard equation of undamped harmonic oscillator (e.g.,
 Bastrukov et al 2009a). The allow for viscosity of stellar material leads to
 exponential damping of amplitude, but the frequency $\nu_A$
 and, hence, the period $P_A$ preserve one and
 the same values as in the case of non-viscous vibrations (Bastrukov et al. 2009c). In
 Fig.1 these latter quantities are plotted as functions of intensity $B$ of undisturbed
 poloidal magnetic field in the neutron star models with indicated mass  $M$ and
 radius $R$.  The practical usefulness of chosen logarithmic scale in this figure is that
 it shows absolute vales of $\nu_A$ and $P_A$ for global vibrations of typical in mass and radius neutron stars.

  In this work, continuing a line of investigations reported in (Bastrukov et al. 2009a, 
  2009b, 2009c, Molodtsova et al. 2010),  we relax the assumption about
  constant-in-time magnetic field and examine the impact of its decay
  on the  vibration energy and period. In section 2, a mathematical background
  of quaking neutron star model is briefly outlined with emphasis on the loss of
  vibration energy caused by depletion of internal magnetic field pressure  and
  resulting vibration-energy powered magneto-dipole radiation.
  The decreasing of magnetic field pressure in the star is presumed to be caused by
  coupling between vibrating star and outgoing material which is expelled by quake,
  but mechanisms of star-envelope interaction resulting in the decay of
  magnetic field, during the time of vibrational relaxation, are not considered.
   It is shown that physically meaningful inferences regarding radiative activity
  of quaking neutron star can be made even when detailed mechanisms of depletion
  of magnetic field pressure in the process of vibrations triggered by quakes are not
  exactly known.
  In section 3, a set of representative examples illustrating the newly disclosed
  effects is considered. The summary of developed theory with emphasis on its
  relevance to electromagnetic activity of quaking magnetars is presented
  in section 4.

 \section{Vibration-energy powered
   magneto-dipole emission of neutron star}

 In what follows we focus on dependence on time of the intensity of poloidal
 magnetic field, $B=B(t)$, about axis of which the star undergoes global torsional
 vibrations and confine our consideration to the model of uniform internal field
 \begin{eqnarray}
  \label{e2.1}
 {\bf B}({\bf r},t)=B(t)\,{\bf b}({\bf r}),\quad  [b_r=\cos\theta,\,\,
 b_\theta=-\sin\theta,\,\, b_\phi=0].
  \end{eqnarray}
  On account of this
  the equation of solid-magnetics, (\ref{e1.2}), takes the form
  \begin{eqnarray}
  \label{e2.2}
 && \rho\, {\ddot {\bf u}}({\bf r},t)=\frac{B^2(t)}{4\pi}
 [\nabla\times[\nabla\times [{\bf u}({\bf r},t)\times {\bf b}({\bf r},t)]]]\times {\bf b}
 ({\bf r}).
  \end{eqnarray}
 Inserting here the following separable form of fluctuating material
 displacements
\begin{eqnarray}
  \label{e2.3}
 && {\bf u}({\bf r},t)={\bf a}({\bf r})\,\alpha(t)
 \end{eqnarray}
 we obtain
  \begin{eqnarray}
  \label{e2.4}
 && \{\rho\, {\bf a}({\bf r})\}\,{\ddot {\alpha}}(t)=2P_B(t)\{
 [\nabla\times[\nabla\times [{\bf a}({\bf r})\times {\bf b}({\bf r})]]]\times {\bf b}
 ({\bf r})\}\,{\alpha}(t),\quad P_B(t)=\frac{B^2(t)}{8\pi}.
  \end{eqnarray}
 Scalar product of (\ref{e2.4}) with the time-independent field of instantaneous
 displacements ${\bf a}({\bf r})$ followed by integration over the star volume
 leads to equation for amplitude $\alpha(t)$ having the form of equation of oscillator
 with depending on time spring constant
 \begin{eqnarray}
  \label{e2.5}
 && {\cal M}{\ddot \alpha}(t)+{\cal K}(t)\alpha(t)=0, \quad {\cal M}=\rho\,m_\ell,
 \quad  {\cal K}(t)=2P_B(t)\, k_\ell,\\
  \label{e2.6}&&  m_\ell=\int \,{\bf a}({\bf r})\cdot {\bf a}({\bf r})\,d{\cal V},\quad
  {\bf a}=A_t\nabla\times [{\bf r}\,r^\ell\,P_\ell(\cos\theta)],\\
  \label{e2.7}
 &&
 k_\ell=\int
 {\bf a}({\bf r})\cdot [{\bf b}({\bf r})\times [\nabla\times[\nabla\times [{\bf a}({\bf
 r})\times {\bf b}({\bf r})]]]]\,d{\cal V}.
 \end{eqnarray}
 The solution of equation of non-isochronal  (non-uniform in
 duration) and non-stationary vibrations with time-dependent frequency [${\ddot
 \alpha}(t)+\omega^2(t)\alpha(t)=0$ where $\omega^2(t)={\cal K}(t)/{\cal M}$] is
 non-trivial and fairly formidable task (e.g.,  Vakman and Vainshtein 1977).
 But solution of such an equation, however, is not a prime purpose of this work. 
 The main subject is the impact of depletion of magnetic-field-pressure on the total 
 energy of Alfv\'en vibrations
   \begin{eqnarray}
 \label{e2.8}
  && E_A(t)=\frac{{\cal M}{\dot \alpha}^2(t)}{2}+\frac{{\cal K}(t)\alpha^2(t)}{2}
  \end{eqnarray}
 and the discrete spectrum of frequency of the toroidal $a$-mode
 \begin{eqnarray}
 \label{e2.9}
 &&\omega_\ell^2(t)=\omega_A^2(t)\,s_\ell^2,\quad
 \omega_A^2(t)=\frac{v_A^2(t)}{R^2},\quad s_\ell^2=\frac{k_\ell}{m_\ell}\,R^2,\\
 \label{e2.10}
 && \omega_\ell^2(t)=B^2(t)\kappa_\ell^2,\quad \kappa_\ell^2=\frac{s_\ell^2}
 {4\pi\rho R^2},\quad s_\ell^2=\left[(\ell^2-1)\frac{2\ell+3}{2\ell-1}\right],\quad
 \ell\geq 2.
  \end{eqnarray}
  It is to be stated clearly from the onset that it is not our goal here to speculate
  about possible mechanisms of neutron star demagnetization and advocate
  conceivable laws of magnetic field decay. The main purpose is  to
  get some insight into the effect of arbitrary law of magnetic field decay in quaking
  neutron star on period of Lorentz-force-driven seismic vibrations and radiative
  activity of the star brought about by such vibrations.
  In the reminder of the paper we remove index $\ell$ in the expression for
  discrete spectrum of frequency of toroidal Alfv\'en mode putting
  $\omega(t)=\omega_\ell(t)$. It seems worth noting that first computation 
  of discrete spectra of frequencies of toroidal Alfv\'en stellar vibrations in standing 
  wave-regime, has been reported by by Chandrasekhar (1956).  The 
  extensive review of other earlier computations of discrete frequency spectra of $a$-modes, $\omega_\ell=\omega_A\,s_\ell$, is given 
 in well-known review of Ledoux and Walraven (1958). In present work, we 
 follow the line of the energy method developed in works (Bastrukov et al 2009a, 2009b, Molodtsova et al 2010) for computing discrete frequencies of node-free
 Alfv\'en vibrations and preserve one and the same notations for frequency spectra 
 as in this latter canonical paper on variable stars.

\subsection{Magnetic-field-decay induced loss of vibration energy}

   The total energy stored in quake-induced Alfv\'en seismic vibrations of the star is
   given by
    \begin{eqnarray}
 \label{e3.1}
  && E_A(t)=\frac{{\cal M}{\dot \alpha}^2(t)}{2}+\frac{{\cal K}(B(t))\alpha^2(t)}
  {2},\quad {\cal K}(B(t))=\omega^2(B(t)){\cal M}.
  \end{eqnarray}
  Perhaps most striking consequence of the magnetic-field-pressure depletion
  during the post-quake vibrational relaxation of neutron star is that it leads
  to the loss of vibration energy at a rate proportional to the rate of magnetic field
  decay
  \begin{eqnarray}
 \label{e3.2}
 \frac{dE_A(t)}{dt}&=&\left\{\dot\alpha(t)[{\cal M}{\ddot \alpha}(t)+{\cal
 K}(B(t))\alpha(t)]+
  \frac{\alpha^2(t)}{2}\frac{d{\cal K}(B(t))}{dt}\right\}= \frac{\alpha^2(t)}
  {2}\frac{d{\cal K}(B)}{dB}\frac{dB(t)}{dt}\\
   \label{e3.3}
  &=&\frac{{\cal M}\alpha^2(t)}{2}\frac{d\omega^2(B)}{dB}\frac{dB(t)}{dt}={\cal
  M}\kappa^2\alpha^2(t)B(t)\frac{B(t)}{dt}.
  \end{eqnarray}
   It is worth emphasizing at this point that the
   magnetic-field-decay induced loss of vibration energy is substantially
   different from the vibration energy dissipation caused by shear viscosity of matter
   resulting in heating of stellar material (Mestel 1999).
   As was noted, the characteristic feature of this latter mechanism
   of vibration energy conversion into the heat (i.e., into the energy of non-coherent
   electromagnetic emission responsible for the formation of photosphere
   of the star)  is that the frequency and, hence, period of vibrations are the same
   as in the case of viscous-free vibrations (Bastrukov et al. 2009c).
   However, it is no longer so in the case under consideration. It follows from
   above that depletion of magnetic field pressure resulting in the loss of total energy
   of Alfv\'en vibrations of the star causes its vibration period to lengthen at a rate
   proportional to the rate of magnetic field decay. The process of magnetic field decay
   depends on many factors of demagnetization of neutron star matter and
   likely primarily on the star-environment communication. Also, it seems
   quite likely that, contrary to viscous dissipation, the loss of vibration energy due to
   decay of magnetic field must be accompanied by coherent (non-thermal)
   electromagnetic radiation. Adhering to this supposition
   in the next sections we consider several representative scenarios of
   magnetic field decay in quaking neutron stars. In so doing,
   special emphasis  is made on conversion of the energy of Lorentz-force-driven
   seismic vibrations into the energy of magneto-dipole emission whose flux
   oscillates with frequency of torsional Alfv\'en magneto-mechanical vibrations of the
   star.

 \subsection{Conversion of vibration energy into power of magneto-dipole
   radiation}

    The point of departure in the study of vibration-energy powered
    magneto-dipole emission of the star (whose radiation power, ${\cal P}$, is given by Larmor's
    formula) is the equation
 \begin{eqnarray}
 \label{e3.4}
  && \frac{dE_A(t)}{dt}={\cal P},\quad {\cal P}=\frac{2}{3c^3}\delta {\ddot {\mbox{\boldmath
  $\mu$}}}^2(t).
  \end{eqnarray}
   Consider a model of quaking neutron star whose torsional magneto-mechanical
  oscillations are accompanied by fluctuations of total magnetic moment preserving
  its initial (in seismically quiescent state) direction: $\mbox{\boldmath
  $\mu$}=\mu\,{\bf n}={\rm constant}$.
  The total magnetic dipole moment should execute oscillations with frequency
  $\omega(t)$ equal to that for magneto-mechanical vibrations of  stellar matter which
  are described by equation for $\alpha(t)$.
  This means that $\delta {\mbox{\boldmath $\mu$}}(t)$ and  $\alpha(t)$
  must be subjected to equations of similar form, namely
  \begin{eqnarray}
   \label{e3.5}
  && \delta {\ddot {\mbox{\boldmath $\mu$}}}(t)+\omega^2(t)
  \delta {\mbox{\boldmath $\mu$}}(t)=0,\\
   \label{e3.6}
  && {\ddot \alpha}(t)+\omega^2(t){\alpha}(t)=0,\quad \omega^2(t)=B^2(t)
  {\kappa}^2.
  \end{eqnarray}
  It is easy to see that equations (\ref{e3.9}) and  (\ref{e3.10}) can be
  reconciled if
  \begin{eqnarray}
   \label{e3.7}
  \delta \mbox{\boldmath $\mu$}(t)=i\,\mbox{\boldmath $\mu$}\,\alpha(t).
  \end{eqnarray}
  Then, from (\ref{e3.5}), it follows
  $\delta {\ddot {\mbox{\boldmath $\mu$}}}=-i\omega^2\mbox{\boldmath $
  \mu$}\alpha$.
  Given this and equating
 \begin{eqnarray}
 \label{e3.8}
  && \frac{dE_A(t)}{dt}={\cal M}\kappa^2\alpha^2(t)B(t)\frac{dB(t)}{dt},
  \end{eqnarray}
  with
  \begin{eqnarray}
   \label{e3.9}
  && {\cal P}=-\frac{2}{3c^3}\mu^2\,\kappa^4 B^4(t){\alpha}^2(t)
  \end{eqnarray}
  we arrive at the equation of  time evolution of magnetic field
   \begin{eqnarray}
  \label{e3.10}
  && \frac{dB(t)}{dt}=-\gamma\,B^3(t),\quad
  \gamma=\frac{2\mu^2\kappa^2}{3{\cal M}c^3}={\rm
  constant}
  \end{eqnarray}
  which yields the following law of its decay
  \begin{eqnarray}
   \label{e3.11}
  B(t)=\frac{B(0)}{\sqrt{1+t/\tau}},\quad \tau^{-1}=2\gamma B^2(0).
  \end{eqnarray}
  The lifetime of magnetic field $\tau$  is regarded as a parameter
  whose value is established from above given relations
  between the period $P$ and its time derivative
  ${\dot P}$ which are  taken from
  observations. And knowing from observations
 $P(t)$ and ${\dot P}(t)$ and estimating $\tau$ one can get information
 about the strength of magnetic field in the star and the magnitude  of its total
 magnetic moment in undisturbed state\footnote{Understandably that
 in the model of vibration-energy powered
  magneto-dipole emission under consideration, the equation
  of magnetic field evolution  is obtained in similar fashion
  as equation for the angular velocity $\Omega$ does in
  the standard model of rotation-energy powered emission of neutron star.
  However, in the model of quaking neutron star vibrating in toroidal
  $a$-mode, the elongation of period is attributed to magnetic field decay, whereas in
  canonical Pacini-Gold model of radio-pulsar the period lengthening is attributed to
  the spin-down of rotating neutron star.}.

 \begin{figure*}[ht]
 \begin{center}
\includegraphics[width=8.cm]{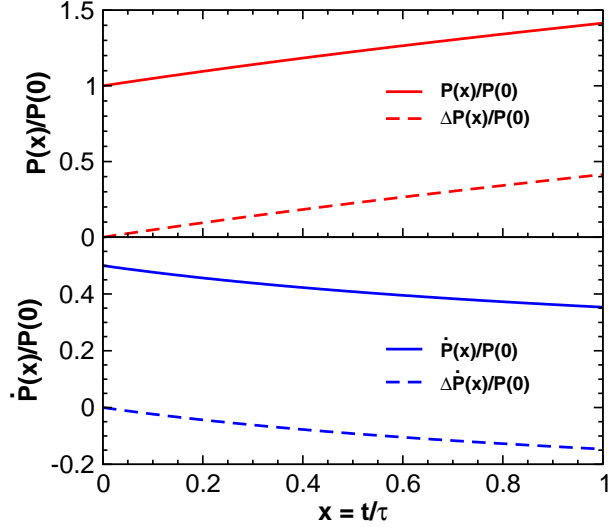}
\end{center}
\caption{
 The upper panel illustrates the magnetic-field-decay induced elongation
 of the period $P(x)$ and difference between periods $\Delta P(x)$ of toroidal $a$-
 mode, defined in the text, plotted as functions of $x=t/\tau$.
 The lower panel shows that in this model of magnetic
 field decay the time derivatives of period ${\dot P}(x)$ and difference between
 periods $\Delta {\dot P}(x)$ are decreasing functions of $x$.}
  \label{fig:p-dot-p1}
 \end{figure*}

\subsection{Lengthening of vibration period}

  The immediate consequence of above line of argument is the
  magnetic-field-decay induced lengthening of vibration period
 \begin{eqnarray}
  \label{e2.11}
  && P(t)=\frac{2\pi}{\omega(t)}=\frac{C}{B(t)},\quad  C=\frac{2\pi}{\kappa}.
  \end{eqnarray}
  The rate of period elongation is given by
  \begin{eqnarray}
  \label{e2.12}
  && \frac{d P(t)}{dt}=-\frac{C}{B^2(t)}\frac{dB(t)}{dt},\quad \quad \frac{dB(t)}{dt}<0.
  \end{eqnarray}
Combining  these equations we obtain the following general relations
 \begin{eqnarray}
  \label{e2.12a}
  && P(t)B(t)={\rm constant},\\
  \label{e2.13}
  && \frac{\dot P(t)}{P(t)}=
  -\frac{\dot B(t)}{B(t)}
  \end{eqnarray}
  which are generic to solely Alfv\'en vibration mode and
  independent of specific form of the magnetic field decay law.
 The difference between periods evaluated at successive moments of time $t_1=0$
 and $t_2=t$ is given by
 \begin{eqnarray}
 \label{e2.14}
 \Delta P(t)=P(t)-P(0)=-P(0)\left[1-\frac{B(0)}{B(t)}\right]>0,\quad B(t)<B(0).
 \end{eqnarray}
 The practical usefulness of these general relations is that they can be
 used as a guide in search for
 fingerprints of Alfv\'en seismic vibrations in data on oscillating emission from
 quaking neutron star.

   In the model under consideration 
 \begin{eqnarray}
 \label{e2.14a}
&&  P(t)=\frac{C}{B(t)},\quad   B(t)=\frac{B(0)}{\sqrt{1+t/\tau}}\\
&& P(0)=\frac{C}{B(0)},\quad   B(0)=B(t=0).
 \end{eqnarray}
   and 
   the time evolution of vibrational period and
   its derivative is described by
    \begin{eqnarray}
  \label{e3.12}
 && P(t)=P(0)\,[1+(t/\tau)]^{1/2},\quad \Delta
 P(t)=P(0)\left[1-\sqrt{1+t/\tau}\right],\\
 \label{e3.13}
 && {\dot P}(t)=\frac{1}{2\tau}\frac{P(0)}{[1+(t/\tau)]^{1/2}}, \quad\to\quad
 {\dot P}(0)=\frac{P(0)}{2\tau}\\
  \label{e3.14}
 && \Delta {\dot P}(t)={\dot P}(t)-{\dot P}(0)=-\frac{P(0)}{2\tau}\left[1-\frac{1}
 {\sqrt{1+t/\tau}}\right].
 \end{eqnarray}
 It follows that lifetime $\tau$ is determined by
 \begin{eqnarray}
 \label{e3.15}
 && {\dot P}(t)\,P(t)=\frac{P^2(0)}{2\tau}={\rm constant}
 \end{eqnarray}
 and the ratio ${\dot P}$ to $P$ is given by
 \begin{eqnarray}
 \label{e3.16}
 &&\frac{{\dot P}(t)}{P(t)}=\frac{1}{2\tau}\,[1+(t/\tau)]^{-1}.
 \end{eqnarray}
 For the sake of illustration of general trends in $P$ and ${\dot P}$, in Fig.2 these
 quantities are plotted as functions of the fraction time $x=t/\tau$ which is ranged in the interval $0<x<1$.
 In Fig. 3, we plot ${\dot P}(t)/{P(0)}$ versus
 ${P(t)}/{P(0)}$.

 \begin{figure*}[ht]
 \begin{center}
 \includegraphics[width=8.cm]{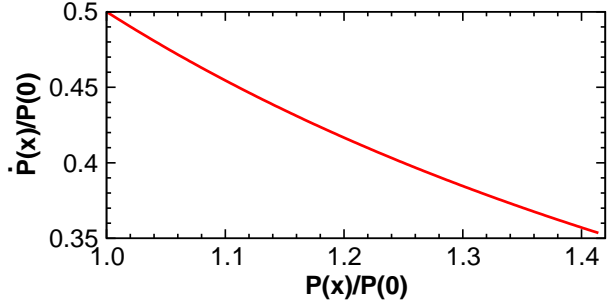}
 \end{center}
  \caption{The interrelation
 between time derivative of period ${\dot P}(x)$ versus period  ${P}(x)$, both
 normalized to period at initial moment of time $P(0)$, computed as functions of
 $x=t/\tau$.}
 \end{figure*}
 
 \subsection{Time evolution of the vibration amplitude}
 
 It is interesting to note that considered model permits 
 exact analytic solution of equation for vibration amplitude $\alpha(t)$ which is convenient to represented as 
\begin{eqnarray}
\label{ee1}
 &&{\ddot \alpha}(t)+\omega^2(t)\alpha(t)=0,\quad \omega^2(t)=\frac{\omega^2(0)}{1+t/\tau},\quad \omega(0)=\omega_A\kappa={\rm constant}.
  \end{eqnarray}
The procedure is as follows. Let us introduce new
variable
\begin{eqnarray}
\label{ee3}
&& s=1+t/\tau,\quad\to\quad t=(s-1)\tau\\
\label{ee4}
&& \frac{\partial s}{\partial t}=\frac{1}{\tau},\quad {\dot \alpha}(t)=\frac{\partial \alpha(t)}{\partial t},\quad \alpha'(s)=\frac{\partial \alpha(s)}{\partial s}
  \end{eqnarray}
so that 
\begin{eqnarray}
 \label{ee6}
 &&{\dot \alpha}(t)=\frac{\partial \alpha(t)}{\partial t}=
 \frac{\partial \alpha(s)}{\partial s}\,\frac{\partial s}{\partial t}=\frac{\alpha'(s)}{\tau},\quad {\ddot \alpha}(t)=\frac{\alpha''(s)}{\tau^2}.
  \end{eqnarray}
In terms of $\alpha(s)$, equation ({\ref{ee1}) takes the form  
\begin{eqnarray}
\label{ee8}
 &&s\alpha''(s)+\beta^2\alpha(s)=0,\quad \beta^2=\omega^2(0)\tau^2={\rm constant}.
  \end{eqnarray}  
 This equation permits exact analytic solution (e.g. Polyanin and Zaitsev, 2004)
\begin{eqnarray}
\label{ee10}
 &&\alpha(s)=s^{1/2}\{C_1\,J_1(2\beta s^{1/2})+C_2Y_1(2\beta s^{1/2})\}
  \end{eqnarray}
 where $J_1(2\beta s^{1/2})$ and $Y_1(2\beta s^{1/2})$ are Bessel functions
 (Abramowitz and Stegun 1972)
\begin{eqnarray}
\label{ee10a}
 J_1(z)=\frac{1}{\pi}\int_0^\pi \cos(z\sin \theta-\theta)d\theta,\quad 
Y_1(z)=\frac{1}{\pi}\int_0^\pi \sin(z\sin \theta-\theta)d\theta,\quad z=2\beta s^{1/2}.
  \end{eqnarray}
The arbitrary constants $C_1$ and $C_2$ can be eliminated from two conditions   
\begin{eqnarray}
\label{ee12}
 &&\alpha(t=0)=\alpha_0,\quad \alpha(t=\tau)=0.
  \end{eqnarray}
where zero-point amplitude  
\begin{eqnarray}
\label{ee12a}
 &&\alpha_0^2=\frac{2{\bar E_A(0)}}{M\omega^2(0)}=\frac{2{\bar E_A(0)}}{K(0)},\quad 
 \omega^2(0)=\frac{K(0)}{M} 
  \end{eqnarray}
is related to the average energy 
${\bar E}_A(0)$ stored in torsional Alfv\'en vibrations at initial (before magnetic field decay) moment of time $t=0$. So, knowing from observations the amplitude of vibrations $\alpha_0$, the average energy ${\bar E}_A(0)$ is established from last equation\footnote{Before magnetic field decay, the star oscillates in 
 the harmonic in time regime, that is, with amplitude $\alpha=\alpha_0\cos(\omega(0) t)$, so that  ${\bar{\alpha^2}}=(1/2)\alpha_0^2$  and ${\bar{\dot \alpha^2}}=(1/2)\omega^2(0)\alpha_0^2$. The average energy stored in 
 such oscillations is given by ${\bar E}_A(0)=(1/2)M{\bar{\dot \alpha^2}}+
(1/2)K(0){\bar{\alpha^2}}=(1/2)M\omega^2(0)\alpha_0^2=(1/2)K(0)\alpha_0^2$
from which equation (\ref{ee12}) is obtained.}.

\begin{figure*}[ht]
 \begin{center}
 \includegraphics[width=12.cm]{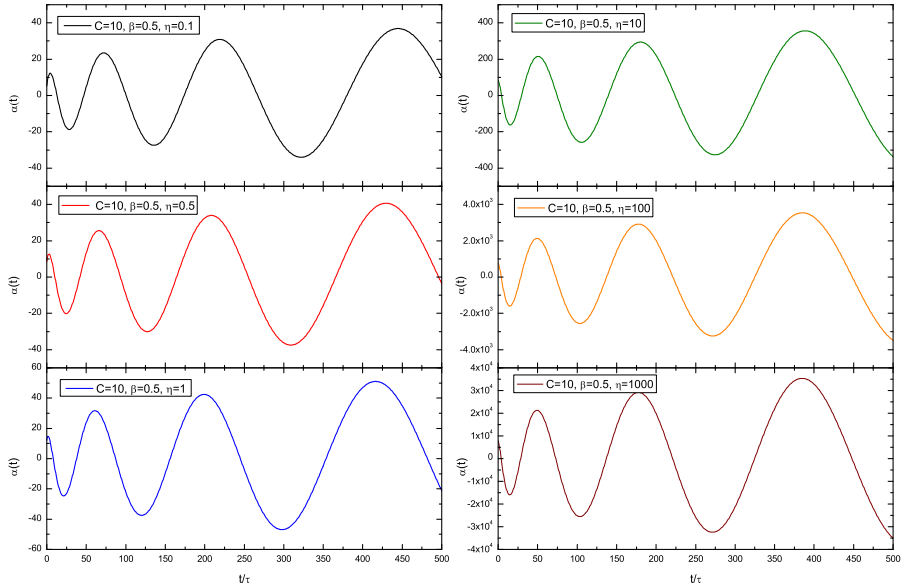}
 \end{center}
 \caption{Vibration amplitude $\alpha(t)$ computed at fixed $C$ and $\beta$ 
 and different values of $\eta$. This shows that variation of this parameter 
 is manifested in change of magnitude of $|\alpha|$, but period elongation 
 is not changed.}
 \end{figure*}
As a result, the general solution of (\ref{ee1}) can be represented in the form
 \begin{eqnarray}
\label{ee11}
 &&\alpha(t)=[1+(t/\tau]^{1/2}C\{J_1(2\beta\,[1+(t/\tau)]^{1/2})-\eta\,Y_1
 (2\beta\,[1+(t/\tau)^{1/2}])\},\\
 \label{ee11a}
 && \eta=\frac{J_1(z(\tau))}{Y_1(z(\tau))},\quad C=\alpha_0[J_1(z(0))-\eta\,Y_1(z(0)]^{-1}.
  \end{eqnarray}

The vibration period lengthening in the process of vibrations is illustrated 
in Fig.4 and Fig.5, where we plot $\alpha(t)$, equation (\ref{ee11}), at different values of parameters $\beta$ and $\eta$ pointed out in the figures. Fig.4 clearly shows that $\eta$ is the parameter regulating magnitude of vibration 
amplitude, the larger $\eta$, the higher is amplitude. However this parameter 
does not affect the rate of period lengthening. As it is clearly seen from Fig.5
both the elongation rate of vibration period and magnitude of vibration 
amplitude are highly sensitive to parameter $\beta$.

\begin{figure*}[ht]
 \begin{center}
 \includegraphics[width=12.cm]{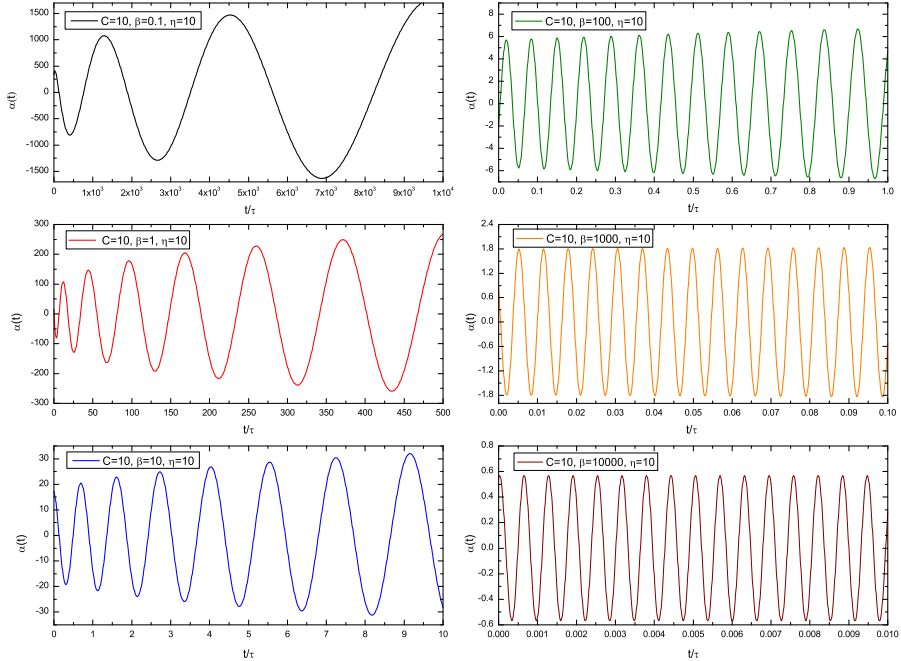}
 \end{center}
 \caption{Vibration amplitude $\alpha(t)$ computed at fixed $C$ and $\eta$
 and different values of $\beta$.}
 \end{figure*}

\section{Representative examples}

 The considered law of magnetic field decay cannot be, of course, regarded
 as universal because it reflects a quite concrete line of argument regarding the
 fluctuations of magnetic moment of the star.
 The interrelation between quake-induced oscillations
 of total magnetic moment  of the star $\delta \mbox{\boldmath $\mu$}(t)$
 and the amplitude ${\alpha}(t)$ of its seismic
 magneto-mechanical oscillations can be consistently interpreted
  with the aid of the function of dipole demagnetization ${\bf f}(B(t))$
 which is defined by the following condition of self-consistency in $\alpha$ of
 right and left hand sides of equation (\ref{e3.4}), namely
 \begin{eqnarray}
  \label{e3.17}
  \delta {\ddot {\mbox{\boldmath $\mu$}}}(t)={\it i}\,{\bf f}(B(t)){\alpha}(t)
  \end{eqnarray}
 The vector-function of dipole demagnetization ${\bf f}(B(t))$  depending on
 decaying magnetic field reflects temporal changes of electromagnetic properties of
 neutron star matter as well as evolution of magnetic-field-promoted coupling
 between neutron star and its environment. This means that specific form
 of this phenomenological function should be motivated by heuristic arguments
 taking into account these factors. With this form of  $\delta {\ddot
 {\mbox{\boldmath $\mu$}}}(t)$, equation (\ref{e3.4}) is  transformed to magnetic
 field decay of the form
  \begin{eqnarray}
 \label{e3.18}
  && \frac{dB(t)}{dt}=-\eta\,\frac{{\bf f}^2(B(t))}{B(t)},\quad
  \eta=\frac{2}{3{\cal M}\kappa^2c^3}={\rm constant}.
  \end{eqnarray}
  In the above considered case this function is given by
   This implies that the function of dipole demagnetization ${\bf f}(B(t))$ has the form
  \begin{eqnarray}
  \label{e3.19}
  && {\bf f}(B(t))=\beta\,B(t)\,{\bf B}(t),\quad \beta= \kappa^2\mu={\rm
  constant}.
  \end{eqnarray}
  As is shown below, the practical
  usefulness of function of dipole demagnetization, ${\bf f}(B(t))$, consists in that it
  provides economic way of studying a vast variety of
  heuristically motivated laws of magnetic field decay, $B=B(t)$,
  whose inferences can ultimately be tested by observations. In the next section,  with no discussion of any
 specific physical mechanism which could be responsible for magnetic field decay,
 we consider a set of representative examples of demagnetization function ${\bf f}
 (B(t))$  some of which lead to the magnetic field decay laws that have
 been regarded before, though in a somewhat different
  context (e.g. Sang, Chanmugam 1987, Srinivasan  et al. 1990,  Goldreich,
  Reisenegger 1992, Urpin et al. 1994, Wang 1997,  Livio et al. 1998,
  Geppert et. al 2001, Bisnovatyi-Kogan, 2002).  In this connection it worth stressing
  that the model under consideration deals with magnetic field decay in the
  course of vibrations triggered by starquake, not with long-term secular decay
  which has been the subject of these latter investigations (see also references
  therein).  In the case in question, a fairly rapid decay of magnetic field during the
  time of post-quake vibrational relaxation of the star is though of as caused, to a
  large extent, by coupling of the vibrating star with material expelled by quake.
  In other words, escaping material removing a part of magnetic flux density from
  the star is considered to be a most plausible reason of depletion of internal
  magnetic field pressure in the star. While physical processes responsible
  for magnetic field decay remain uncertain, practical significance of above,
  admittedly heuristic, line of argument is that it leads to meaningful conclusion
  (regarding elongation of periods of oscillating magneto-dipole emission) even
  when detailed mechanisms of depletion of magnetic field pressure in the course of
  vibrations are not exactly known.

\begin{figure*}[ht]
\begin{center}
\includegraphics[width=7.5cm]{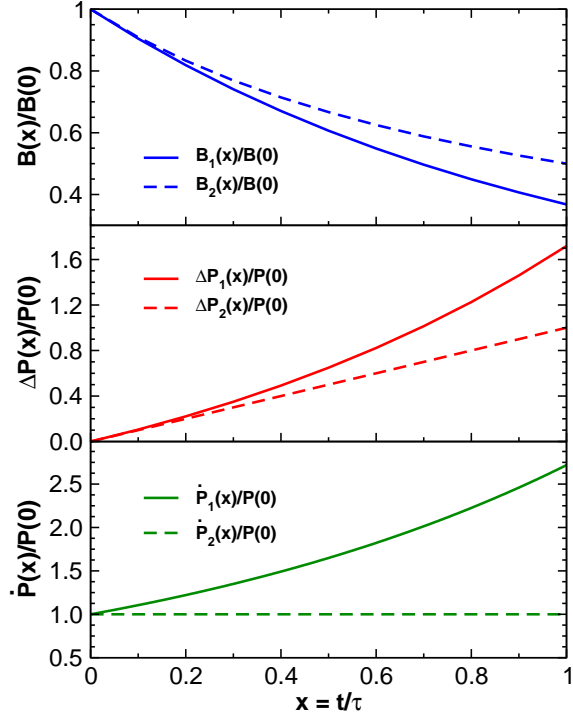}
\end{center}
 \caption{
 The upper panel -- the period $P(x)$ and difference between periods $\Delta P(x)$ of
 toroidal $a$-mode as functions of $x=t/\tau$. The middle panel
  shows that in this model of
 magnetic field decay the time derivatives of period ${\dot P}(x)$ and difference
 between periods $\Delta {\dot P}(x)$ are decreasing functions of $x$. The lower
 panel, the caused by field decay rate of the $a$-modes vibration period lengthening,
 normalized to period of initial moment of time, caused  by magnetic field decay with
 time evolution of considered in the text representative examples
 $i=1$ and 2.}

\end{figure*}
\begin{figure*}[ht]
 \begin{center}
 \includegraphics[width=7.5cm]{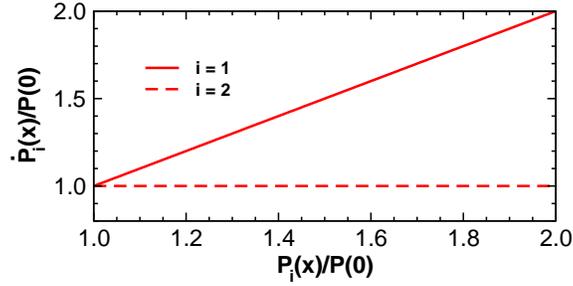}
 \end{center}
 \caption{The rate of period versus period of toroidal $a$-mode
  computed in 1-st and 2-nd representative models.}
 \end{figure*}

 1. As a first representative example, consider a model of quaking neutron star
 in which function of dipole
 demagnetization is given by
 \begin{eqnarray}
 \label{ea.2}
 {\bf f}(B(t))=\beta {\bf B}(t).
 \end{eqnarray}
 This leads to exponential decay of magnetic filed
 \begin{eqnarray}
 \label{ea.3}
 \frac{dB(t)}{dt}&=&-\frac{B(t)}{\tau},\quad\to\quad B(t)=B(0)\,e^{-t/\tau},\quad
 \tau^{-1}=\eta\beta^2
 \end{eqnarray}
 where $B(0)$ is the intensity of magnetic field before quake.
 The period and its derivative are given by
 \begin{eqnarray}
 \label{ea.4}
 P(t)=P(0)\,e^{t/\tau},\quad {\dot P}(t)=\frac{P(0)}{\tau}\,e^{t/\tau}.
 \end{eqnarray}
 The difference between periods, evaluated at the moment of time $t_2=t$ and
 $t_1=0$ is given by
 \begin{eqnarray}
 \label{ea.6}
 \Delta P(t)=P(t)-P(0)=P(0)\,e^{t/\tau_1}(1-e^{-t/\tau}).
 \end{eqnarray}
 In the initial stage of decay, when $(t/\tau)<<1$, the last equation is reduced to
 $\Delta P(t)\approx P(0)\,(t/\tau)\,[1+(t/\tau)]$. The parameter of magnetic field
 lifetime is given by
 \begin{eqnarray}
 \label{ea.5}
 \tau^{-1}=\frac{{\dot P}(t)}{P(t)}.
 \end{eqnarray}
 In Fig.6 the above period and its derivative are plotted as functions of fraction
 time $x=t/\tau$.

 2.  Consider a neutron star model whose function of demagnetization is given by
 \begin{eqnarray}
 \label{ea.7}
 {\bf f}(B(t))=\beta\sqrt{B(t)}\, {\bf B}(t).
 \end{eqnarray}
 The resultant equation of magnetic field evolution reads
 \begin{eqnarray}
 \label{ea.8}
 \frac{dB(t)}{dt}&=&-\gamma\,B^{2}(t),\quad \gamma=\eta\beta^2
 \end{eqnarray}
 and thus implying that rate of magnetic field decay is proportional
 to the density of magnetic field energy stored in the star. In this case one has
 \begin{eqnarray}
 \label{ea.9}
 B(t)=B(0)\left(1+\frac{t}{\tau}\right)^{-1}.
 \end{eqnarray}
 The elongation of vibration period and a rate of its lengthening
 are given by
 \begin{eqnarray}
 \label{ea.10}
 P(t)=P(0)\left(1+\frac{t}{\tau}\right), \quad
 {\dot P}(t)=\frac{P(0)}{\tau}={\rm constant}.
 \end{eqnarray}
 The difference between periods is given by $\Delta P(t)=P(0)({t}/{\tau})$.
 The decay time is eliminated from the ratio
 \footnote{Similar analysis can be performed
 for the demagnetization function of the form
 \begin{eqnarray}
 \nonumber
 {\bf f}(B(t))=\beta\sqrt{B^{m-1}(t)}\, {\bf B}(t), \quad m =3,4,5...
 \end{eqnarray}
 which leads to
 \begin{eqnarray}
 \nonumber
 &&\frac{dB(t)}{dt}=-\gamma\,B^{m}(t),\quad \gamma=\eta\beta^2,\\
 \nonumber
 && B(t)=\frac{B(0)}{[1+t/\tau_m]^{1/(m-1)}},\quad \tau_m^{-1}=\gamma
 (m-1)B^{m-1}(0).
 \end{eqnarray}
 }
 \begin{eqnarray}
 \label{ea.11}
 \frac{{\dot P}(t)}{P(t)}={\tau}^{-1}\left(1+\frac{t}{\tau}\right)^{-1}\quad
 {\tau}^{-1}=\frac{{\dot P}(0)}{P(0)}.
 \end{eqnarray}
 The inferences of this model regarding period and its derivative as functions of time
 are presented in Fig. 7.

 \begin{figure*}[ht]
 \begin{center}
 \includegraphics[width=8.cm]{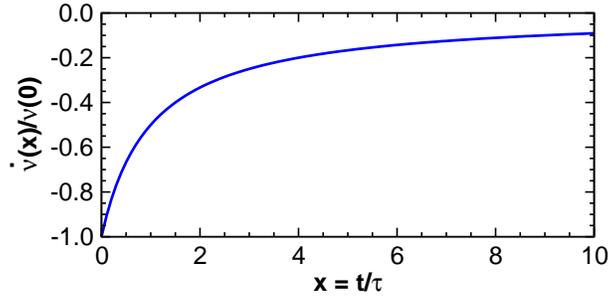}
  \caption{
 The rate of frequency  ${\dot \nu}(x)$ normalized to $\nu(0)$ as a function of
 $x=t/\tau$ for logarithmic law of magnetic filed decay.}
 \end{center}
 \end{figure*}

 3. Finally, consider a model with quite sophisticated
 function of dipole demagnetization
 \begin{eqnarray}
 \label{e4.9}
 && {\bf f}(B(t))=\sqrt{\frac{B(0)}{\tau+t}}\frac{{\bf B}(t)}{\sqrt{\eta B(t)}}
  \end{eqnarray}
 which is interesting in that it leads to equation
  \begin{eqnarray}
 \label{e4.10}
 &&  \frac{dB(t)}{dt}=-\frac{B(0)}{\tau}\left(1+\frac{t}
 {\tau}\right)^{-1}=-\frac{B(0)}{\tau+t}
 \end{eqnarray}
 yielding a fairly non-trivial logarithmic law of magnetic field decay
  \begin{eqnarray}
 \label{e4.11}
  &&  B(t)=B(0)\left[1-\ln\left(1+\frac{t}{\tau}\right)\right].
 \end{eqnarray}
 The corresponding frequency of toroidal $a$-mode as a function of time
 is plotted in Fig.8 and it may be interesting to note that similar shape of ${\dot \nu}
 (t)/\nu(0)$ exhibit data on post-glitch emission of PSR J1846-0258 (Livingstone et al.
 2010).

 In this model
\begin{eqnarray}
 \label{e4.12}
&&  P(t)=P(0)\left[1-\ln\left(1+\frac{t}{\tau}\right)\right]^{-1},\\
 \label{e4.13}
 &&  {\dot P}(t)=\frac{dP(t)}{dt}=\frac{P(0)}{\tau+t}\left[1-\ln\left(1+\frac{t}
 {\tau}\right)\right]^{-2},\\
 \label{e4.14}
 && \tau^{-1}=\frac{{\dot P}(0)}{P(0)}.
 \end{eqnarray}
 These examples show that while the period elongation is the common effect
 of depletion of  magnetic field pressure, the interrelations between periods and its
 derivatives substantially depend on specific form of the magnetic field decay law.

\section{Summary}

 The main purpose of this work was to examine inferences of asteroseismic model of
 neutron star undergoing quake-induced torsional Alfv\'en
 vibrations about axis of its dipole magnetic moment which are
 accompanied by monotonic depletion of internal magnetic field pressure.
 It is shown this latter process results in the loss of vibration energy
 and changes the character of temporal evolution of vibrations from harmonic in time
 vibrations (with constant frequency) in static, time-independent, magnetic field to
 non-isochronal vibrations with monotonically
 decreasing frequency in decaying magnetic field. The presented analysis
 suggests that the loss of vibration energy caused by depletion of magnetic
 field pressure should be accompanied by coherent (non-thermal)
 oscillating magneto-dipole radiation (with frequency of luminosity oscillations equal
 to the frequency of Alfv\'en seismic vibrations of the star).  In this context it is
 appropriate to remind a seminal work of Hoyle, Narlikar and Wheeler (1964) in which it has been pointed out for the first time that vibrating neutron star should
 operate like Hertzian magnetic dipole deriving radiative power of magneto-dipole
 emission from the energy of magneto-mechanical vibrations (see, also, Pacini 2008).
 What is newly disclosed here is that conversion of vibration
  energy into the energy of magneto-dipole electromagnetic radiation can be realized
  when, and only when, torsional Alfv\'en vibrations are accompanied by decay of
  magnetic field.

 All above suggests that considered for the first time theory of vibration-energy powered emission of neutron star 
 is relevant to electromagnetic activity of magnetars - neutron stars endowed
 with magnetic field of extremely high intensity the
 radiative activity of which is ultimately related to the magnetic field decay  (Duncan \& Thompson 1992, Thompson \& Duncan 1995).
 This subclass of highly magnetized compact objects is commonly associated with soft gamma repeaters and anomalous X-ray pulsars
  (e.g., Harding 1999, Kouveliotou 1999, Woods and Thompson 2006,  Mereghetti 2008)
  --  young  isolated and seismically active
 neutron stars (Chen et al. 1996, Franco et al. 2000). The magnetar quakes
 are exhibited by short-duration thermonuclear
 gamma-ray flash followed by rapidly oscillating X-ray flare of
 several-hundred-seconds duration. During this latter stage of quake-induced 
 radiation  of magnetar, a long-periodic
 (5-12 sec) modulation of brightness was observed lasting for about 3-4 minutes
 (e.g., Harding 1999, Kouveliotou 1999). Taking into account such long-periodic and
 rapidly dying variability of magnetar pulsating radiation is the feature
 which is non-typical for young neutron stars powered by energy of rotation, it has
 been suggested (Bastrukov et al. 2002)  that detected long-periodic pulsating emission is powered by  the energy of torsional magneto-elastic vibrations of magnetar triggered by quake.  In view of key role of ultra-strong magnetic field it is quite likely that quasi-periodic oscillations (QPOs) of outburst flux from SGR 1806-20 and SGR 1900+14 discovered in observations reported in (Israel et al. 2005, Watts \&  Strohmayer 2006, Terasawa et al. 2006) are produced
 by torsional seismic vibrations predominately sustained by Lorentz force
 (Bastrukov et al 2009b, 2009c) which are accompanied, as was argued above, by
 monotonic decay of background magnetic field. If so, the predicted elongation
 of QPOs period  of oscillating outburst emission from quaking magnetars
 should be traced in existing and future observations.

  The authors are indebted to colleagues from Physics Department and
  Kavli Institute of Astronomy and Astrophysics, Peking University, Beijing,
  for valuable discussions and suggestions. 
  Special thanks go to Alexey Kudryashov (Saratov University, Russia) who paid 
  our attention to the existing of exact solution of obtained equation for vibration 
  amplitude.


\begin{thebibliography}

\bibitem{AS-64} Abramowitz M., Stegun I.,  1972,  Handbook of mathematical functions (Dover)

\bibitem{BP-97} Bastrukov S. I., Molodtsova I. V., Papoyan V. V., Podgainyi D. V. 1997, Ap, 40, 46

\bibitem{B-99b} Bastrukov S. I., Molodtsova I. V., Podgainy D. V., Weber F., Papoyan V. V., 1999, PPN, 30, 436

\bibitem{B-07a} Bastrukov S. I., Chang H.-K., Mi\c sicu \c S., Molodtsova I. V.,  Podgainy D. V., 2007a, Int. J. Mod. Phys. A, 22, 3261

\bibitem{B-07b} Bastrukov S. I., Chang H.-K., Takata J., Chen G.-T., Molodtsova I. V., 2007b, MNRAS, 382, 849


\bibitem{B-09a} Bastrukov S. I., Chen K.-T., Chang H.-K., Molodtsova, I. V., Podgainy D. V., 2009a, ApJ, 690, 998

\bibitem{B-09b} Bastrukov S. I., Chang H.-K., Molodtsova I.V., Wu E.-H., Chen K.-T., Lan S.-H., 2009b, ApSS, 323, 235

\bibitem{B-09c} Bastrukov S. I., Molodtsova I. V., Chang H.-K., Takata J., Xu R.X., 2009c, arXiv0910.3048B, in progress

\bibitem{B-02} Bastrukov S., Yang J., Kim M., Podgainy D., 2002,
 in Current high-energy emission around black holes, eds. Lee  C.-H., Chang  H.-Y.
  (World Scientific) p.334


\bibitem{BK-02} Bisnovatyi-Kogan G. S., 2002, MmSAI, 73, 318


\bibitem{Car-86} Carroll B. W., Zweibel E. G., Hansen C. J., McDermott P. N., Savedoff M. P., Thomas J. H., van Horn, H. M., 1986, ApJ, 305, 767


\bibitem{Chandra-56} Chandrasekhar S., 1956, ApJ, 124, 571

\bibitem{Cheng-96} Cheng B., Epstein R. I., Guyer R. A., Young A. C., 1996,  Nature, 382, 518

\bibitem{TD-92}  Duncan R. C., Thompson C., 1992, ApJ, 392, L9

\bibitem{TD-96} Duncan R. C., Thompson C.,  1996, ApJ, 473, 322

\bibitem{G-01} Geppert U., Page D., Colpi M., Zannias T., 2000, ASPC 202, 681

\bibitem{GR-92}  Goldreich P., Reisenegger A., 1992, ApJ, 395, 250

\bibitem{ISR-05} Israel G. L. et al., 2005, ApJ, 628, L53

\bibitem{K-99} Kouveliotou C.,  1999, PNAS, 96, 5351


\bibitem{Lee-07} Lee U., 2007, MNRAS, 374, 1015

\bibitem{Lee-08} Lee U., 2008, MNRAS, 385, 2069

\bibitem{LW-58} Ledoux P., Walraven T., 1958, HDP, 51, 353


\bibitem{L-10} Livingstone M. A.,  Kaspi V. M., Gavriil F. P., 2010, ApJ,  710, 1710

\bibitem{LXF-98} Livio M., Xu  C., Frank, J., 1998, ApJ, 492, 298

\bibitem{H-99} Harding A. K., 1999, AN, 320, 260

\bibitem{H-09} Heyvaerts J., Bonazzola S., Bejger M., Haensel P., 2009, AA, 496, 317


\bibitem{HNW-64} Hoyle F.,  Narlikar J. V., Wheeler J. A., 1964, Nature 203, 914

\bibitem{FLE-00} Franco L. M., Link B., Epstein R. I., 2000,  ApJ, 543, 987

\bibitem{M-99} Mestel L.,  1999,  Stellar Magnetism (Clarendon)

\bibitem{M-08} Mereghetti S., 2008, AARv, 15, 225

\bibitem{M-10} Molodtsova I. V., Bastrukov S. I., Chen K.-T., Chang H.-K., 2010, ApSS, 327, 1

\bibitem{MOON-85} Moon F. C., 1984,  Magneto-Solid Mechanics  (Wiley)

\bibitem{P-08}  Pacini F., 2008, Proc. MEASRIM No1, Hady A. and Wanas M. I. eds, 75

\bibitem{PZ-04} Polyanin  A. D., Zaitsev V. F., 2004, Handbook of Nonlinear Partial Differential Equations (Chapman and Hall)  

\bibitem{RR-03} Rincon F., Rieutord M., 2003, A\&A, 398, 663

 \bibitem{CS-87} Sang Y., Chanmugam G., 1987, ApJ, 323, L6

  \bibitem{SE-09} Shaisultanov R., Eichler D., 2009, ApJ, 702, L23

 \bibitem{S-09} Sotani H., Kokkotas K. D., Stergioulas N., 2009, JPhCS, 189, 2038

\bibitem{SBMT-90}  Srinivasan G., Bhachataria D., Muslimov A.G., Tsygan A.L., 1990, CSci, 59, 31

\bibitem{Tokyo-SGR} Terasawa T., et al. 2006, JPhCS, 31, 76


\bibitem{VV-77}  Vakman D. E., Vainshtein L. A., 1977, SvPhU, 20, 1002


\bibitem{W-97} Wang J. C. L., 1997, ApJ, 486, L119

\bibitem{WS-06} Watts A. L., Strohmayer, T. E., 2006, ApJ, 637, L117

\bibitem{WT-06} Woods P. M., Thompson, C.  2006, in Compact Stellar X-ray Sources, eds. Lewin W., van der Klis M. (Cambridge University Press)

\bibitem{UCS-94} Urpin V. A., Chanmugam G., Sang Y.,  1994, ApJ 433, 780

\bibitem{Xu-03} Xu R. X.,  2003,  ApJ, 596, L59

\bibitem{Xu-09} Xu R. X.,  2009,  J. Phys.  G 36, 064010


\end{thebibliography}
\end{document}